\documentclass[%
 reprint,
 amsmath,amssymb,
 aps,
]{revtex4-2}

\usepackage{graphicx}
\usepackage{dcolumn}
\usepackage{bm}
\usepackage{subfigure}

\begin{document}

\preprint{APS/123-QED}

\title{Electron Tunneling Spectroscopy of the anisotropic Kitaev Quantum Spin Liquid Sandwiched with Superconductors}

\author{Shi-Qing Jia}
\affiliation{Key Laboratory of Materials Physics, Institute of Solid State Physics, HFIPS, Chinese Academy of
                          Sciences, Hefei 230031, China}
\affiliation{Science Island Branch of Graduate School, University of Science and Technology of China,
                                  Hefei 230026, China}

\author{Ya-Min Quan}
\affiliation{Key Laboratory of Materials Physics, Institute of Solid State Physics, HFIPS, Chinese Academy of
                          Sciences, Hefei 230031, China}
\author{Liang-Jian Zou}
 \email{zou@theory.issp.ac.cn}
\affiliation{Key Laboratory of Materials Physics, Institute of Solid State Physics, HFIPS, Chinese Academy of
                           Sciences, Hefei 230031, China}
\affiliation{Science Island Branch of Graduate School, University of Science and Technology of China,
                                  Hefei 230026, China}

\author{Hai-Qing Lin}
 \email{haiqing0@csrc.ac.cn}
\affiliation{Beijing Computational Science Research Center, Beijing 100193, China 
}
\affiliation{Department of Physics, Beijing Normal University, Beijing 100875, China
}


\date{\today}

\begin{abstract}
We present the electron tunneling transport and spectroscopic characters of a superconducting {\it Josephson} junction with a barrier of single anisotropic Kitaev quantum spin liquid (QSL) layer. We find that the dynamical spin correlation features are well reflected in the direct-current differential conductance $dI^{c}/dV$ of the single-particle tunneling, including the unique spin gap and dressed itinerant Majorana dispersive band, in addition to an energy shift $2\Delta$ of two-lead superconducting gaps. From the spectral characters, we identify different topological quantum phases of the anisotropic Kitaev QSL. We also present the zero-voltage {\it Josephson} current $I^{s}$ which displays residual features of the anisotropic Kitaev QSL. These results pave a new way to measure the dynamical spinon or Majorana fermion spectroscopy of the Kitaev and other spin liquid materials.

\begin{description}
\item[PACS numbers] 75.10.Kt, 75.10.Jm, 74.50.+r
\end{description}
\end{abstract}

\maketitle



\section{\label{sec:level1} Introduction}

The quantum spin liquid (QSL) phase, which consists of various spin singlet pairings in the spin structure without breaking any constituent symmetries of their underlying lattice, has attracted great attention \cite{Balents2010,Meng2010}. Enormous efforts have been made to understand the essence of the QSLs, and earlier studies focused on the geometrically and magnetic frustrated interaction \cite{Anderson1973,Fazes1973}. However, the essence and unique characters of the QSL states remain great debates \cite{Mezzacapo2012,Yu2014}.
More than a decade ago Kitaev proposed an exactly solvable model on the two-dimension (2D) honeycomb lattice \cite{Jackeli2009}, which shows that the interaction frustration drives a ground state of gapless or gapped $Z_{2}$ QSL with fractionalized excitations \cite{Kitaev2006}. The QSL state with gapped excitations has the Abelian anyons \cite{Kitaev2003}, the one with gapless excitations may have the non-Abelian anyon excitations \cite{Read1989}. Due to topological protection and large degeneracy of these anyons, the Majorana fermion excitations and its braiding group in the gapless QSL state were expected to be applicable for the quantum computing storage and quantum computation \cite{Kitaev2006,Hegde2019}.
However, how to excite and detect the dynamics of these Majorana fermion modes in Kitaev systems remains unknown.

On the other hand, the {\it Josephson} tunneling junctions, which are constructed of two superconducting (SC) leads separated by an insulating or metallic barrier, provide a well probe to measure the quasi-particle information of the central region through the quantum tunneling transport \cite{Bakurskiy2019,Xiang2007}. A great deal of central materials, such as insulators \cite{Kleinsasser1994}, normal metals \cite{Morpurgo1997}, quantum dots \cite{SunWW2000,Zhu652001,Sun2002}, ferromagnets \cite{Demler1997,Ryazanov2001,Gingrich2014} and antiferromagnets \cite{Gorkov2002,Bulaevskii2017} have been studied. In order to explore the exotic spin correlations and fractional excitations of the Majorana fermions through the transports of single electrons and Cooper pairs, especially the inelastic spin scattering process \cite{Carrega2020,Konig2020,Feldmeier2020}, it is worth constructing novel SC-Kitaev layer-SC tunneling junctions to reveal its current dynamics associated with exotic spin excitations in Kitaev layer.
In realistic candidate materials for the Kitaev layer, the spin interactions are usually anisotropic\cite{Yamaji2016,Choi2012,Banerjee2016,Jia2021}, thus we employ anisotropic Kitaev layer in the designed SC Josephson junctions.

In this {\it paper}, we utilize the current and conductance features of the SC-anisotropic Kitaev layer-SC tunneling junctions to characterize the dynamical spin correlations of the central-zone Kitaev materials. We adopt the non-equilibrium Green's function \cite{Zhu652001} and the few-particle response method \cite{Knolle2014,Knolle2016} to obtain the formulae of the single-particle and {\it Josephson} tunneling currents.
We find that the dynamical spin susceptibility explicitly displays in the direct current (DC) single-particle differential conductance spectrum $dI^{c}/dV$, and from its spectral features, we could confirm the different topological quantum phases of the anisotropic Kitaev QSL.
One expects that the SC-anisotropic Kitaev QSL-SC mesoscopic hybrid systems with weak links may open a fruitful research field, not only because of the abundant fundamental features from the interplay between Kitaev physics and SC, but also of the potential application for design and development of new quantum devices.

\section{\label{sec:level2} Model and Tunneling of the SC-Kitaev QSL-SC Junction}

\subsection{\label{sec:level21} The SC-Kitaev QSL-SC junction and tunneling process}

\begin{figure}[htbp]
	\centering
	\includegraphics[angle=0, width=0.95 \columnwidth]{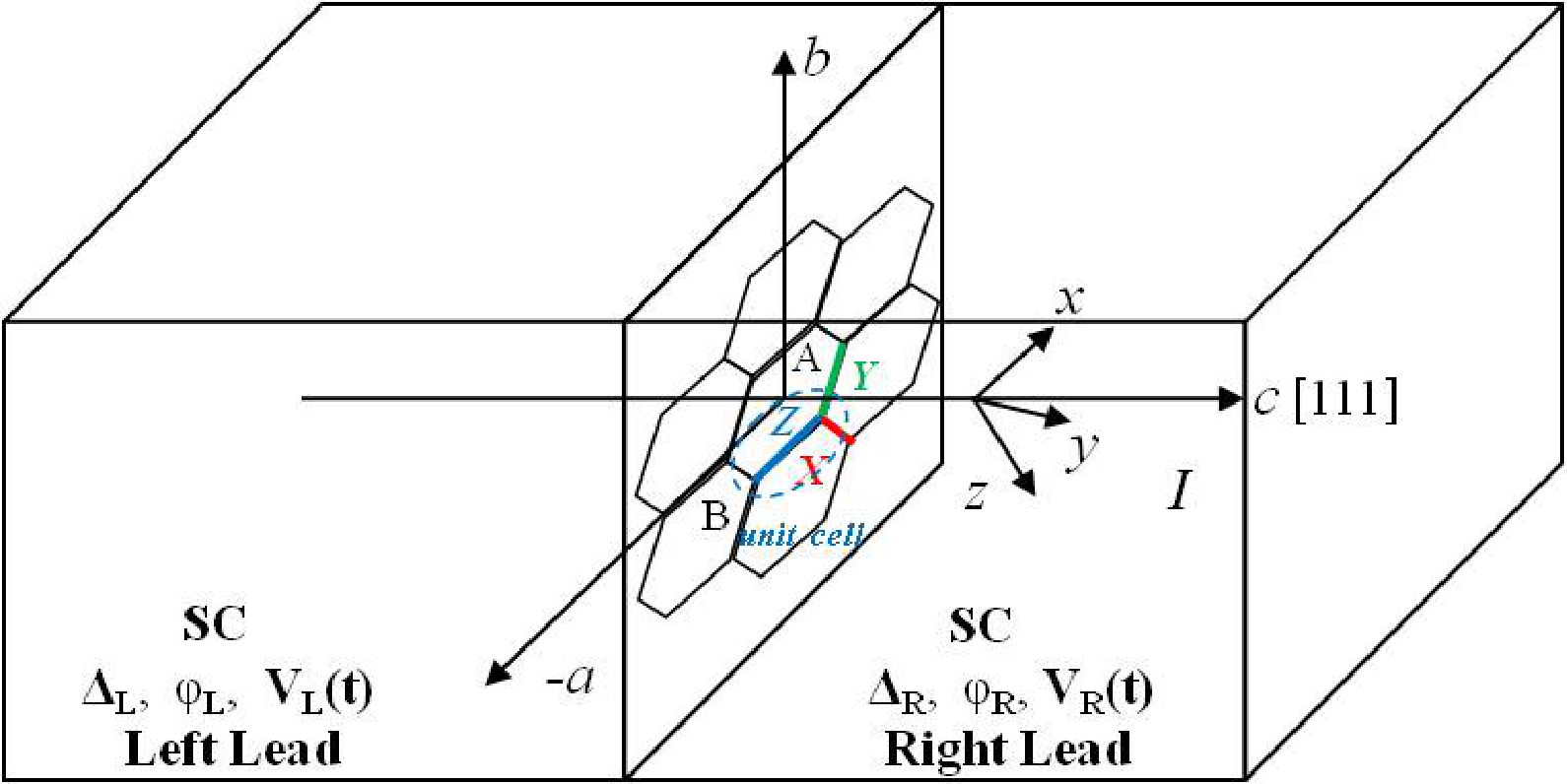}
	\caption{(Color online) Schematic superconductor-Kitaev QSL-superconductor tunneling junction. The left (right) side is the SC lead with gap $\Delta_{L}$ ($\Delta_{R}$), phase $\phi_{L}$ ($\phi_{R}$) and electric potential $V_{L}(t)$ ($V_{R}(t)$). The central region is a single-layer Kitaev material in the $ab$ plane.}
	\label{fig:state1}
\end{figure}

\begin{figure}[htbp]
\centering
\includegraphics[angle=0, width=1.0 \columnwidth]{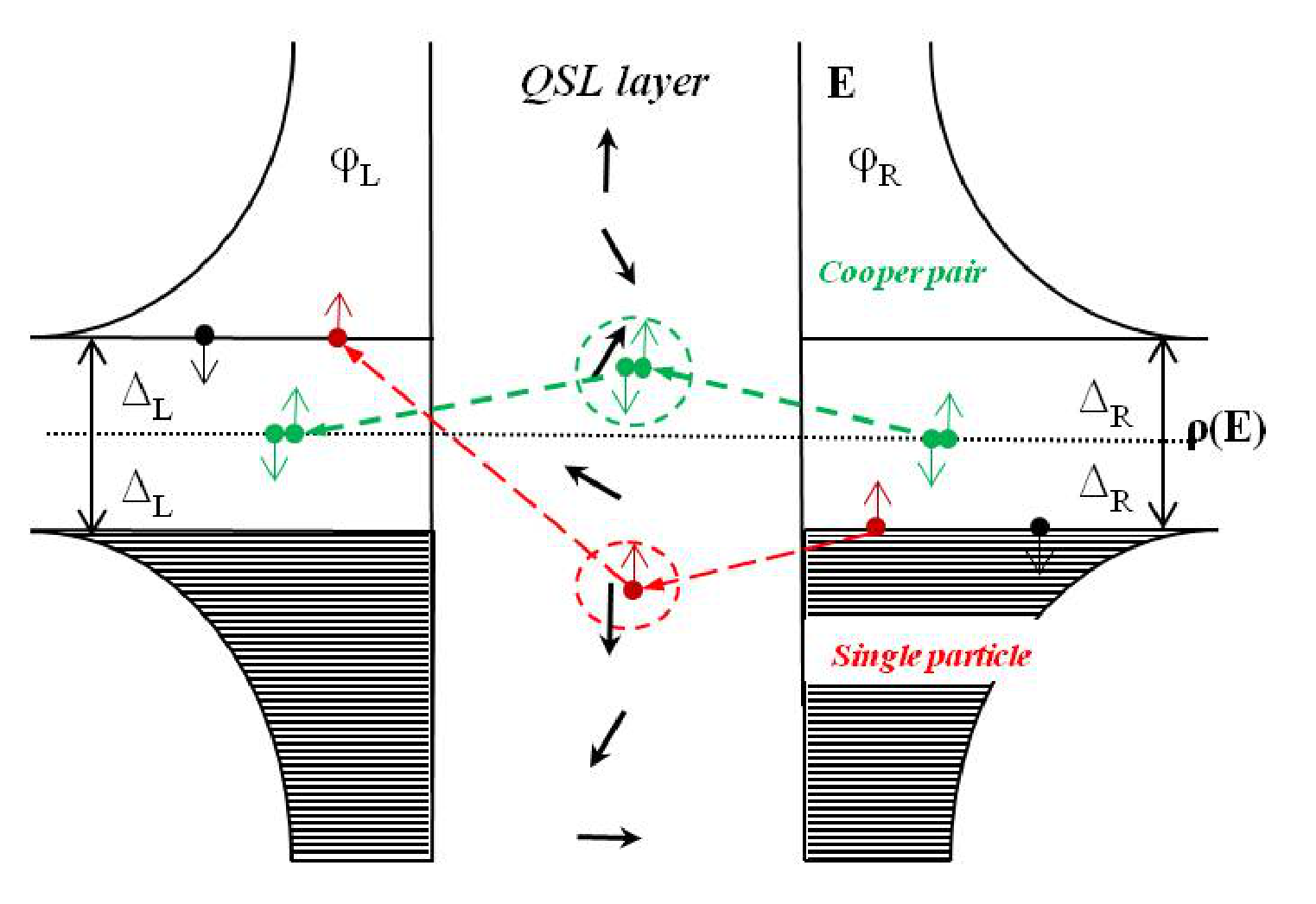}
\caption{(Color online) Sketched diagram of the single-particle (red) and Cooper pair (green) tunneling processes
in the superconductor-Kitaev QSL-superconductor {\it Josephson} junction. The left and right sides are the
bare density-of-states (DOS) distributions $\rho(E)$ of the two SC leads, and the center is the Kitaev
QSL layer. The circles indicate the $s-d$ exchange processes of single particle and a Cooper
pair with local spin, respectively.}
\label{fig:state2}
\end{figure}

We construct a Kitaev Josephson junction, where a single-layer Kitaev insulator is the barrier, sandwiched by two leads consisting of two conventional s-wave superconductors. Here the SC leads may be Nb, or Pb metals, or their alloys NbTi and Nb$_{3}$Sn, {\it etc.}, and the central Kitaev layer may be $\alpha$-RuCl$_{3}$ or Na$_2$IrO$_{3}$ single layer, which are the candidate materials of the Kitaev QSL \cite{Banerjee2016}.
Such a SC-Kitaev QSL-SC {\it Josephson} junction is shown in Fig.~\ref{fig:state1}. Since the Kitaev material is a kind of transition-metal Mott insulator with strong electronic correlation, the tunneling of conduction electrons between left and right SC leads is scattered by the local spins in the central region, as shown in Fig.~\ref{fig:state2}. The scattering strength is $s-d$-type exchange coupling $J$.

For this set-up, the tunneling current consists of normal single-particle one and Josephson one. We can describe the normal single-particle tunneling process as follows: firstly, the electrons at the bottom of SC gap in the right lead enter the Kitaev layer, and occupy the high energy levels to form the virtual double occupied states. The propagation of the electrons would be modulated by the dynamical spin susceptibility of the Kitaev QSL in the spin-conserving channel, as well as in the spin-flipping process with spin fluctuations. Finally, the electrons leave the Kitaev layer with constant or opposite spins and go to the top of the SC gap in the left SC lead.

Moreover, the tunneling process of the SC Cooper pairs can be addressed as follows: the Cooper pair in the right lead firstly tunnels into the central Kitaev region, splitting as the quasi-electron and quasi-hole with opposite spins. Afterwards, the quasi-electron and quasi-hole would go through the similar virtual transitions as the single particles with the modulation of the Kitaev QSL. Once tunneling out of the central Kitaev region, the separated quasi-electrons and quasi-holes would recombine to SC Cooper pairs. These tunneling processes of single particles and Cooper pairs could be qualitatively described by the sketched diagram shown in Fig.~\ref{fig:state2}.

\subsection{\label{sec:level22} Model Hamiltonian and Formulae}

The total Hamiltonian of the SC-Kitaev QSL-SC tunneling junction shown in Fig.~\ref{fig:state1} and Fig.~\ref{fig:state2} consists of three parts as follows: the left and right SC electrodes $H_{Lead,n} (n = L, R)$, the single-layer Kitaev material in the central scattering region $H_{cen}$, and the $s-d$ exchange interaction part between the SC leads and central material $H_{T}$. So $H=\sum_{n=L,R}H_{Lead,n}+H_{cen}+H_{T}$, and

\begin{eqnarray}
\label{eq:Hamiltonian1}
&&H_{Lead,n} = \sum_{k\sigma}\epsilon^{0}_{nk\sigma}a^{\dag}_{nk\sigma}a^{}_{nk\sigma} +
       \sum_{k}\Delta_{n}\left[a^{}_{n,-k\downarrow}a^{}_{nk\uparrow} + h.c. \right],  \nonumber\\
&&H_{cen} = -K_{X}\sum_{\langle ij\rangle_{X}}\hat{\sigma}^{x}_{i}\hat{\sigma}^{x}_{j}
     - K_{Y}\sum_{\langle ij\rangle_{Y}}\hat{\sigma}^{y}_{i}\hat{\sigma}^{y}_{j}
     - K_{Z}\sum_{\langle ij\rangle_{Z}}\hat{\sigma}^{z}_{i}\hat{\sigma}^{z}_{j},  \\
&&H_{T} = - \sum_{i}\left\{\frac{1}{2}\tilde{J}_{i}(t)\left[ \begin{array}{c} \hat{\sigma}^{z}_{i}
       \left(a^{\dag}_{Li\uparrow}a^{}_{Ri\uparrow} - a^{\dag}_{Li\downarrow}a^{}_{Ri\downarrow}\right) \\
     + \hat{\sigma}^{+}_{i}a^{\dag}_{Li\downarrow}a^{}_{Ri\uparrow}
     + \hat{\sigma}^{-}_{i}a^{\dag}_{Li\uparrow}a^{}_{Ri\downarrow} \end{array} \right] + h.c.\right\}, \nonumber
\end{eqnarray}
where $a^{\dag}_{nk\sigma}$ and $c^{\dag}_{i\sigma}$ are the creation operators of electrons in the SC leads and Kitaev layer, respectively, and $a^{\dag}_{ni\sigma}$ is the Fourier transform of $a^{\dag}_{nk\sigma}$ on the $i$th site of the 2D interface between the SC leads and Kitaev layer. $\hat{\sigma}_{i}^{x(y,z)}=\sum_{\sigma\sigma'}c^{\dag}_{i\sigma} \sigma_{\sigma\sigma'}^{x(y,z)}c^{}_{i\sigma'}$ are the twice spin components, $\hat{\sigma}^{\pm}_{i}=\hat{\sigma}^{x}_{i}\pm i\hat{\sigma}^{y}_{i}$, and $\sigma_{\sigma\sigma'}^{x(y,z)}$ are the Pauli matrices. Let the two SC leads be the $s-wave$ superconductors and their order parameters $\tilde{\Delta}_{n}=\Delta_{n}e^{-i\phi_{n}}$ with magnitudes $\Delta_{n}$ and phases $\phi_{n}$. $\epsilon^{0}_{nk\sigma}$ is the single-electron energy. $K_{X}$, $K_{Y}$ and $K_{Z}$ are the spin coupling constants along the $X$, $Y$ and $Z$ bonds in the central Kitaev layer, and they satisfy the conditions $K_{X}=K_{Y}>0$ and $K_{X}+K_{Y}+K_{Z}=3K$ for the anisotropic Kitaev model. $J_i$ is the $s-d$ exchange matrix element between the electrons in the SC leads and the local spins in Kitaev layer.
In the presence of external electric potential $V_{n}(t)(n=L,R)$, the exchange parameter becomes voltage dependence of $\tilde{J}_{i}(t)=J_{i}\exp[i(\phi_{L}-\phi_{R})-(i/\hbar)\int_{0}^{t}e(V_L(t_1)-V_R(t_1))dt_1]$ through a unitary transformation, leaving only the perturbation term $H_T$ explicitly depends on time \cite{SunWW2000}.

The tunneling current from the left SC lead to the central region reads,
\begin{eqnarray}
\label{eq:Hamiltonian2}
\lefteqn{I_{}(t) = -e\left\langle\frac{dN_{L}(t)}{dt}\right\rangle =
     \frac{ie}{\hbar}\left\langle\left[N_{L}(t),H(t)\right]\right\rangle} \\
 & &=-\frac{e}{\hbar}\operatorname{Re}\sum_{i}\tilde{J}_{i}(t)i\left\langle \begin{array}{c}
     \hat{\sigma}^{z}_{i}\left(a^{\dag}_{Li\uparrow}a^{}_{Ri\uparrow}
     - a^{\dag}_{Li\downarrow}a^{}_{Ri\downarrow}\right) \\
     {}+ \hat{\sigma}^{+}_{i}a^{\dag}_{Li\downarrow}a^{}_{Ri\uparrow}
     + \hat{\sigma}^{-}_{i}a^{\dag}_{Li\uparrow}a^{}_{Ri\downarrow} \end{array} \right\rangle.\nonumber
\end{eqnarray}
It actually contains two parts: the normal single-particle tunneling current and SC {\it Josephson} current, and both of them stem from the inelastic scattering with the spin-conserving ($m=zz$) and spin-flipping ($m=xx,yy$) processes,
\begin{eqnarray}
\label{eq:Hamiltonian3}
\lefteqn{I_{}(t) = -\frac{2e}{\hbar}\operatorname{Re}\sum_{ij,m}\int_{-\infty}^{t} \frac{dt_{1}}{\hbar}J_{i}J_{j}} \\
 & &  \left\{ \begin{array}{c} e^{\frac{ieV(t-t_{1})}{\hbar}}
      \left[ \begin{array}{c} \tilde{g}^{r}_{m,LR,ij}\left(t,t_{1}\right)G^{<}_{m,ji}\left(t_{1},t\right) \\
   {}+\tilde{g}^{<}_{m,LR,ij}\left(t,t_{1}\right)G^{a}_{m,ji}\left(t_{1},t\right) \end{array} \right]  \\
   {}+e^{\frac{ieV(t+t_{1})}{\hbar}}e^{i\phi}\left[ \begin{array}{c}
      \tilde{g}'^{r}_{m,LR,ij}\left(t,t_{1}\right)G^{<}_{m,ji}\left(t_{1},t\right) \\
   {}+\tilde{g}'^{<}_{m,LR,ij}\left(t,t_{1}\right)G^{a}_{m,ji}\left(t_{1},t\right) \end{array} \right]
      \end{array} \right\}.\nonumber
\end{eqnarray}

Throughout this paper we only consider the DC voltage $V=V_{L}-V_{R}$ and $\phi=\phi_{L}-\phi_{R}$ is the phase difference between the left and right SC leads. Define $G(g)^{r,a,<}_{m,ji}(t_1,t)$ with superscripts $^r$, $^a$, and $^<$ as the dressed (bare) retarded, advanced, and lesser Green's functions of spin correlation in the central region, respectively. $\tilde{g}^{r,a,<}_{m,LR,ij}(t,t_1)$ and $\tilde{g}'^{r,a,<}_{m,LR,ij}(t,t_1)$ are bare normal and anomalous Green's functions of electron-hole modes and Cooper pairs between left and right SC leads, respectively. For example, the advanced Green's functions can be written as follows:
\begin{eqnarray}
\label{eq:Hamiltonian4}
g^{r}_{m,ji}(t_1,t)&=&-i\theta(t_1-t)
\langle[0.5\hat{\sigma}^{\alpha}_{j}(t_1),0.5\hat{\sigma}^{\alpha}_{i}(t)]\rangle,\\
\tilde{g}^{r}_{m,LR,ij}(t,t_1)&=&-i\theta(t-t_1) \nonumber\\
&&\langle[\hat{\sigma}^{\alpha}_{\sigma\sigma'}a^{\dag}_{Li\sigma}a^{}_{Ri\sigma'}(t),
\hat{\sigma}^{\alpha}_{\sigma\sigma'}a^{\dag}_{Rj\sigma}a^{}_{Lj\sigma'}(t_1)]\rangle,  \nonumber\\
\tilde{g}'^{r}_{m,LR,ij}(t,t_1)&=&-i\theta(t-t_1) \nonumber\\
&&\langle[\hat{\sigma}^{\alpha}_{\sigma\sigma'}a^{\dag}_{Li\sigma}a^{}_{Ri\sigma'}(t),
\hat{\sigma}^{\alpha}_{\sigma\sigma'}a^{\dag}_{Lj\sigma}a^{}_{Rj\sigma'}(t_1)]\rangle, \nonumber
\end{eqnarray}
where $m=\alpha\alpha,\alpha=x,y,z$. The details are shown in Sec. A of {\it Supplementary Materials} \cite{Supple2020}.

With zero bias voltage, we have only DC {\it Josephson} current $I^{s}$ generated by the tunneling of Cooper electron pairs through the Kitaev QSL. Moreover, at $V\neq0$, we are much interested at the DC current $I^{c}$ and its conductance $dI^{c}/dV$ of the normal single-particle tunneling. Thus, the DC single-particle and {\it Josephson} current terms in the first-order approximation can be obtained as follows:
\begin{eqnarray}
\label{eq:Hamiltonian5}
 I^{c}_{} &=& \frac{4e}{\hbar}\sum_{ij,m}\int\frac{d\epsilon}{2\pi}J_{i}J_{j}\,\operatorname{Im}\left[\tilde{g}^{r}_{m,LR,ij}
                  \left(eV-\epsilon\right)\right]  \nonumber\\
              & & \operatorname{Im}\left[g^{r}_{m,ij}\left(\epsilon\right)\right]\left[n\left(\epsilon\right)
                  - n\left(\epsilon - eV\right)\right],  \\
 I^{s}_{} &=& \frac{4e}{\hbar}\sum_{ij,m}\int\frac{d\epsilon}{2\pi}J_{i}J_{j}\,
                  \operatorname{Im}\left[\tilde{g}'^{r}_{m,LR,ij}\left(\epsilon\right)g^{r}_{m,ji}\left(\epsilon\right)\right] \nonumber\\
              & & n\left(\epsilon\right)\sin\phi, \nonumber
\end{eqnarray}
respectively, where $n(\epsilon)=1/[\exp(\epsilon/k_{B}T)-1]$ is the Bose-Einstein distribution function. As seen in Eq.~(\ref{eq:Hamiltonian5}), $I^{c}$ obviously depends on the dynamical spin susceptibility $S^{m}_{ij}(\epsilon)=-2\operatorname{Im}[g^{r}_{m,ij}(\epsilon)]$ of the Kitaev QSL, the spectral weight of electron-hole modes $C^{m}_{LR,ij}(\epsilon)=-2\operatorname{Im}[\tilde{g}^{r}_{m,LR,ij}(\epsilon)]$ between the two SC leads and the occupation difference between spins and electron-hole modes. Similarly, $I^{s}$ is weighted by the hybridization spectrum of spins and Cooper pairs $A^{m}_{hy,ij}(\epsilon)=2\operatorname{Im}[\tilde{g}'^{r}_{m,LR,ij}(\epsilon)g^{r}_{m,ji}(\epsilon)]$ and the Bose-Einstein occupation $n(\epsilon)$. In these inelastic scattering processes, the electron-hole modes or Cooper pairs with charge between left and right SC leads transfer energy to the central spin system \cite{Konig2020}.

Actually, further analysis reveals that both the normal and anomalous Green's functions of the two leads have the same $zz$, $xx$ and $yy$ components because of the time-reversal symmetry. We have $\tilde{g}^{r}_{m,LR,ij}(\epsilon)$=$\tilde{g}^{r}_{0,LR,ij}(\epsilon)$ and $\tilde{g}'^{r}_{m,LR,ij}(\epsilon)=\tilde{g}'^{r}_{0,LR,ij}(\epsilon)$ for $m=xx,yy$, and $zz$, respectively.
At the same time, the unique feature of QSL leads to that $g^{r}_{m,ji}(\epsilon)$ is a short-range spin correlation in real space and only the on-site and nearest-neighbour (NN) ones are nonzero, which is explained later in Sec.~\ref{sec:level23}. So the currents have two part contributions from the on-site and NN $X$, $Y$, $Z$ bonds.
Therefore, we can simplify the tunneling currents $I^{c}$ and $I^{s}$ at zero temperature as
\begin{widetext}
\begin{eqnarray}
\label{eq:Hamiltonian6}
 I^{c}_{} &=& \frac{8e}{\hbar}NJ^{2}\sum_{m}\int_{0}^{eV} \frac{d\epsilon}{2\pi}
                  \left\{ \begin{array}{c} \operatorname{Im}\left[\tilde{g}^{r}_{0,LR,AA}\left(eV - \epsilon\right)\right]
                   \operatorname{Im}\left[g^{r}_{m,AA}\left(\epsilon\right)\right] \\
               {}+\sum_{\langle AB\rangle}\operatorname{Im}\left[\tilde{g}^{r}_{0,LR,BA}\left(eV - \epsilon\right)\right]
                   \operatorname{Im}\left[g^{r}_{m,BA}\left(\epsilon\right)\right] \end{array} \right\},  \nonumber\\
I^{s}_{} &=& \frac{8e}{\hbar}NJ^{2}\sum_{m}\int_{0}^{\infty} \frac{d\epsilon}{2\pi}\sin\phi
                 \left\{ \begin{array}{c} \operatorname{Im}\left[\tilde{g}'^{r}_{0,LR,AA}\left(\epsilon\right)
                  g^{r}_{m,AA}\left(\epsilon\right)\right] \\
               {}+\sum_{\langle AB\rangle}\operatorname{Im}\left[\tilde{g}'^{r}_{0,LR,AB}\left(\epsilon\right)
                  g^{r}_{m,BA}\left(\epsilon\right)\right] \end{array} \right\}.
\end{eqnarray}
\end{widetext}
Here the indexes of the sublattices, AA and AB, stand for the on-site and NN configurations, and $J_{i}=J$ for each site $i$. $N$ is the number of unit cell of honeycomb lattice.

Then, the normal and anomalous two-body Green's functions can be evaluated through the frequency summations over the combinations of left- and right-lead single-body Green's functions, in the $4\times4$ Nambu representation $(a_{nk\uparrow}^{}\, {a}_{n,-k\downarrow}^{\dagger}\, a_{nk\downarrow}^{}\, {a}_{n,-k\uparrow}^{\dagger})$. The details can be seen in Sec. B of {\it Supplementary Materials} \cite{Supple2020}. We thus obtain that
\begin{subequations}
\label{eq:Hamiltonian7}
\begin{eqnarray}
\label{eq:Hamiltonian7a}
\lefteqn{\tilde{g}_{0,LR,AA(BA)}^{r}\left(\epsilon\right)=\frac{s^{2}}{2}\int{\frac{d^{2}k}{4\pi^{2}}}
\displaystyle\int{\frac{d^{2}p}{4\pi^{2}}}{e^{i\mathbf{\left(k+p\right)\cdot R_{AA(BA)}}}}} \nonumber\\
&&\left\{\frac{1}
{\epsilon-{E_{Rp}}-{E_{Lk}}+i{0^{+}}}
-\frac{1}{\epsilon +{E_{Rp}}+{E_{Lk}}+i{0^{+}}}\right\}
\end{eqnarray}
\begin{eqnarray}
\label{eq:Hamiltonian7b}
\lefteqn{{\tilde{g}}'^{r}_{0,LR,AA(AB)}\left(\epsilon\right)=-\frac{s^{2}}{2}\int{\frac{d^{2}k}{4\pi ^{2}}}
\displaystyle\int{\frac{d^{2}p}{4\pi^{2}}}{e^{i\mathbf{\left(k+p\right)\cdot R_{AA(AB)}}}}} \\
&&\frac{\Delta_{L}\Delta_{R}}{E_{Lk}E_{Rp}}
\left\{\frac{1}{\epsilon-{E_{Rp}}-{E_{Lk}}+i{0^{+}}} - \frac{1}{\epsilon+{E_{Rp}}+{E_{Lk}}+i{0^{+}}}\right\}\nonumber
\end{eqnarray}
\end{subequations}
Here $E_{nk(p)}=\sqrt{\epsilon_{nk(p)}^{2}+\Delta_{n}^{2}}$, and the parabolic energy dispersions $\epsilon_{Lk}=\hbar^{2}k^{2}/2m^{*}-E_{F}$, $\epsilon_{Rp}=\hbar ^{2}p^{2}/2m^{*}-E_{F}$. $m^{*}$ is the effective mass of electron, $E_{F}$ is the Fermi energy level, and set $\hbar=1$. $s$ is the area of unit cell of SC-Kitaev layer-SC interface in the SC leads. $R_{AA}=0$ and $R_{AB}=X,Y,Z$ for the on-site and NN ones, respectively.

Assuming that $k_{F}=1/a_{s}$ and $E_{F}=20K$, where $k_{F}$ and $a_{s}$ are the Fermi wave vector and lattice constant of two SC leads. Since the exchanged momenta between the SC leads and Kitaev layer are constrained by $0 \leq |\mathbf{q}|\leq 2k_{F}$, the product $\mathbf{q\cdot X(Y,Z)}$ ($\mathbf{q=k+p}$) can be taken to zero for simplicity \cite{Carrega2020,Konig2020} in the Green's functions with the NN contribution. This is suitable for the ``bad metal'' like Nb or Pb with the small Fermi wave vectors. Hence, in the leads, we have $\tilde{g}^{r}_{0,LR,AB(BA)}(\epsilon)\approx\tilde{g}^{r}_{0,LR,AA(BB)}(\epsilon)$ and $\tilde{g}'^{r}_{0,LR,AB(BA)}(\epsilon)\approx\tilde{g}'^{r}_{0,LR,AA(BB)}(\epsilon)$. Then the imaginary part of the normal retarded Green's function of the two SC leads can be further simplified as follows,
\begin{subequations}
\label{eq:Hamiltonian8}
\begin{eqnarray}
\label{eq:Hamiltonian8a}
 \lefteqn{\operatorname{Im}\left[\tilde{g}^{r}_{0,LR,AA(BA)}\left(\epsilon\right)\right]=-2\pi\rho_L\rho_R} \\
 & &\left\{ \begin{array} {cc}
 \displaystyle\int_{\Delta_L}^{\epsilon}{dE\frac{E}{\sqrt{E^{2}-\Delta_L^2}}}\frac{\left(\epsilon-E\right)}{\sqrt{\left( \epsilon-E\right)^2-\Delta_R^2}},\, \epsilon \ge E+\Delta_R \\
 \displaystyle\int_{\Delta_L}^{-\epsilon}{dE\frac{E}{\sqrt{E^2-\Delta_L^2}}}\frac{\left(\epsilon+E\right)}{\sqrt{\left( \epsilon+E\right)^2-\Delta_R^2}},\, \epsilon \le -E-\Delta_R
 \end{array} \right. ,\nonumber
 \end{eqnarray}
 as well as the imaginary and real parts of the anomalous retarded Green's function
 \begin{eqnarray}
 \label{eq:Hamiltonian8b}
 \lefteqn{\operatorname{Im}\left[\tilde{g}'^{r}_{0,LR,AA(AB)}\left(\epsilon\right)\right]=\frac{\pi}{2}\rho_L\rho_R} \\
 & &\left\{ \begin{array} {cc}
 \displaystyle\int_{\Delta_L}^{\epsilon}{dE\frac{\Delta_L}{\sqrt{E^{2}-\Delta_L^2}}}\frac{\Delta_R}{\sqrt{\left( \epsilon-E\right)^2-\Delta_R^2}},\, \epsilon \ge E+\Delta_R \\
 \displaystyle\int_{\Delta_L}^{-\epsilon}{dE\frac{\Delta_L}{\sqrt{E^2-\Delta_L^2}}}\frac{-\Delta_R}{\sqrt{\left( \epsilon+E\right)^2-\Delta_R^2}},\epsilon \le-E-\Delta_R
 \end{array} \right., \nonumber\\
 & &\mbox{Re}\left[\tilde{g}'^{r}_{0,LR,AA(AB)}\left(\epsilon\right)\right]=\int_{-\infty}^{\infty}
 {\frac{d\omega}{2\pi}\frac{\left(-2\right)\mbox{Im}\left[\tilde{g}'^{r}_{0,LR,AA(AB)}\left(\omega\right)\right]}
 {\epsilon - \omega}}. \nonumber
\end{eqnarray}
\end{subequations}
Here we calculate the real part of Green's function by the Kramers-Kronig transformation. The normal density of states (DOS) in the 2D interface $\rho_{L(R)}=m^{*}a_{s}^{2}/2\pi\hbar^{2}$. More details can be seen in Sec. B of {\it Supplementary Materials} \cite{Supple2020}.

Therefore, we can obtain the DC single-particle differential conductance $dI^{c}/dV$ and the derivative of the DC {\it Josephson} current $I^{s}$ with respect to $\Delta$, $dI^{s}/d\Delta$,  as
\begin{eqnarray}
\label{eq:Hamiltonian9}
\frac{dI^{c}}{dV} &=& \frac{2e^{2}}{\hbar}NJ^{2}\sum_{m}\int_{0}^{eV} \frac{d\epsilon}{2\pi}
                  \left\{ \frac{d[C^{0}_{LR,AA}\left(eV - \epsilon\right)]}{dV}
                  S^{m}\left(\epsilon\right) \right\},  \nonumber\\
\frac{dI^{s}}{d\Delta} &=& \frac{4e}{\hbar}NJ^{2}\sum_{m}\int_{0}^{\infty} \frac{d\epsilon}{2\pi}
                  \frac{d[A^{m}_{hy}\left(\epsilon\right)]}{d\Delta}\sin\phi.
\end{eqnarray}
Here the total dynamical spin susceptibility, the total hybridization spectrum of spins and Cooper pairs, and the equally weighted spectrum of electron-hole modes are defined as
\begin{eqnarray}
\label{eq:Hamiltonian10}
S^{m}(\epsilon)&=&-2\operatorname{Im}[g^{r}_{m}(\epsilon)],  \nonumber\\
A^{m}_{hy}(\epsilon)&=&2\operatorname{Im}\{\tilde{g}'^{r}_{0,LR,AA}(\epsilon)g^{r}_{m}(\epsilon)\}, \nonumber\\
C^{0}_{LR,AA}(\epsilon)&=&-2\operatorname{Im}[\tilde{g}^{r}_{0,LR,AA}(\epsilon)],
\end{eqnarray}
respectively, where the total Green's function of spin correlation $g^{r}_{m}(\epsilon)=g^{r}_{m,AA}(\epsilon)+\Sigma_{\langle AB\rangle}g^{r}_{m,BA}(\epsilon)$.
Once obtaining the Green's functions $\tilde{g}^{r}_{0,LR,AA}(\epsilon)$, $\tilde{g}'^{r}_{0,LR,AA}(\epsilon)$ and $g^{r}_{m}(\epsilon)$, we could get the DC single-particle current and its differential conductance numerically, as well as the zero-voltage {\it Josephson} current at zero temperature.

\subsection{\label{sec:level23} Dynamics of the Kitaev model}

Next, we need the total Green's function of spin correlation of anisotropic Kitaev QSL, $g^{r}_{m}(\epsilon)$, whose imaginary part corresponds to the dynamical spin susceptibility, $S^{m}(\epsilon)$. We would evaluate the $S^{m}(\epsilon)$ by employing the few-particle-response method and $g^{r}_{m}(\epsilon)$ via the Kramers-Kronig transformation.

The Kitaev model $H_{cen}$ in Eq.~(\ref{eq:Hamiltonian1}) can be exactly solved by introducing four Majorana fermions $b_{i}^{\alpha}$ ($\alpha=x,y,z$) and $c_{i}$ per site for the local spins, {\it i.e.} $\hat{\sigma}_{i}^{\alpha}=ic_{i}b_{i}^{\alpha}$. Define the bond operators $\hat{u}_{ij}^{\alpha}
=ib_{i}^{\alpha}b_{j}^{\alpha}$ on the NN bond $\langle ij\rangle_{\Lambda}$ ($\Lambda=X,Y,Z$), respectively.
Their eigenvalues are $u_{ij}^{\alpha}=\pm1$, and they commute with $H_{cen}$ and with each other. So the Kitaev model can be expressed in terms of the different sets of $\{u_{ij}^{\alpha}\}$ and the Majorana fermions \cite{Kitaev2006,Knolle2016},
 \begin{eqnarray}
 \label{eq:Hamiltonian11}
  H_{cen}=i\sum_{\Lambda,\langle ij\rangle_{\Lambda}}K_{\Lambda}
  u_{ij}^{\alpha}c_{i}c_{j},
 \end{eqnarray}
where the product of all bond operators around a plaquette, $W_{p}=\prod_{i,j\in p}u_{ij}^{\alpha}\doteq \pm1$, can define the flux sectors. The eigenstates of this model are $Z_{2}$ gauge fluxes threading the plaquettes and Majorana fermions (or spinons) propagating between sites in this $Z_{2}$ gauge field \cite{Knolle2016}. And their wave vectors $|\Phi\rangle$ are the direct product of bond (gauge flux) and Majorana-matter-fermion degrees of freedoms, $|\Phi\rangle=|F\rangle \otimes |M\rangle$. The ground state is within the zero-flux sector with $W_{p}=1$ ($u_{ij}^{\alpha}$=1) for all plaquettes.

\begin{figure*}[htbp]
\centering
\subfigure[]{
\label{fig:state3a}
\includegraphics[angle=0, width=0.66 \columnwidth]{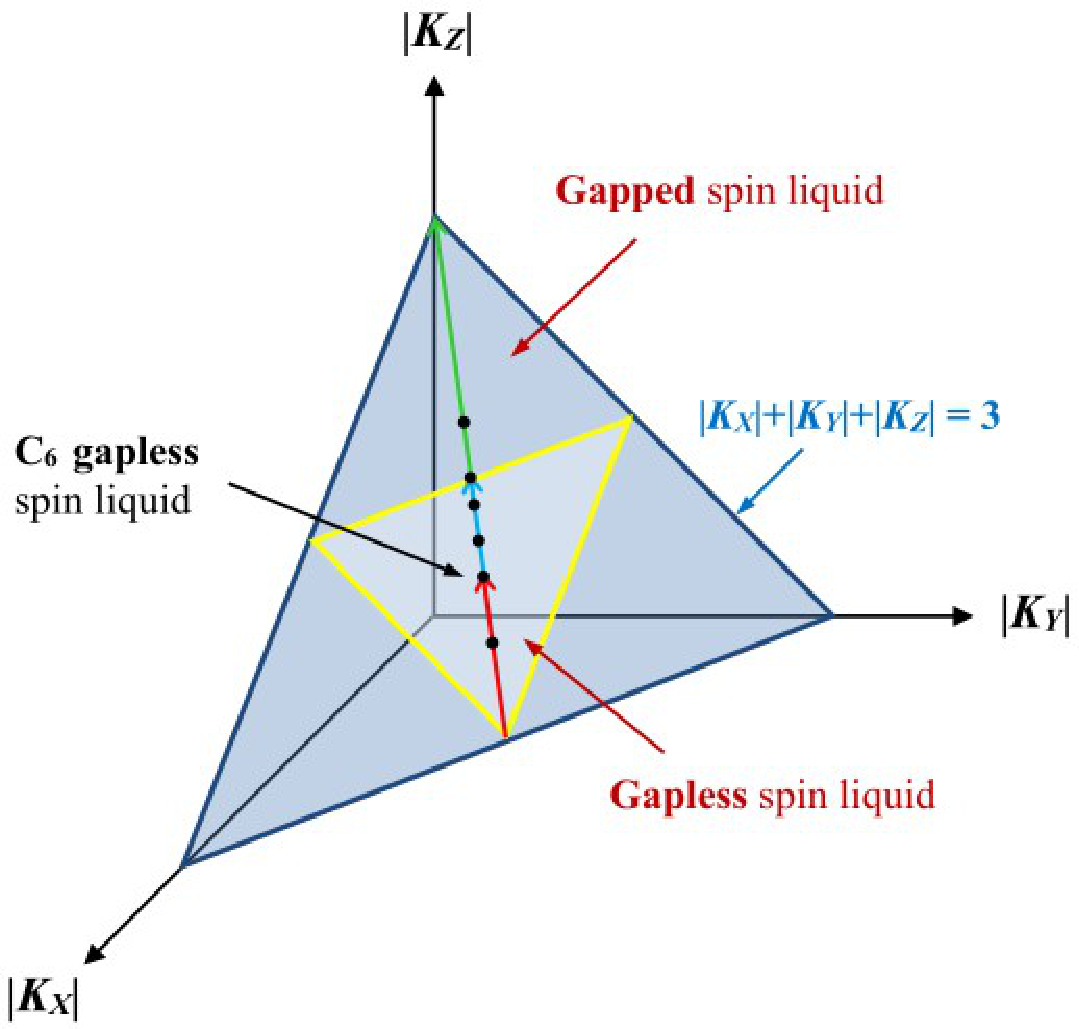}}
\hspace{0in}
\subfigure[]{
\label{fig:state3b}
\includegraphics[angle=0, width=0.66 \columnwidth]{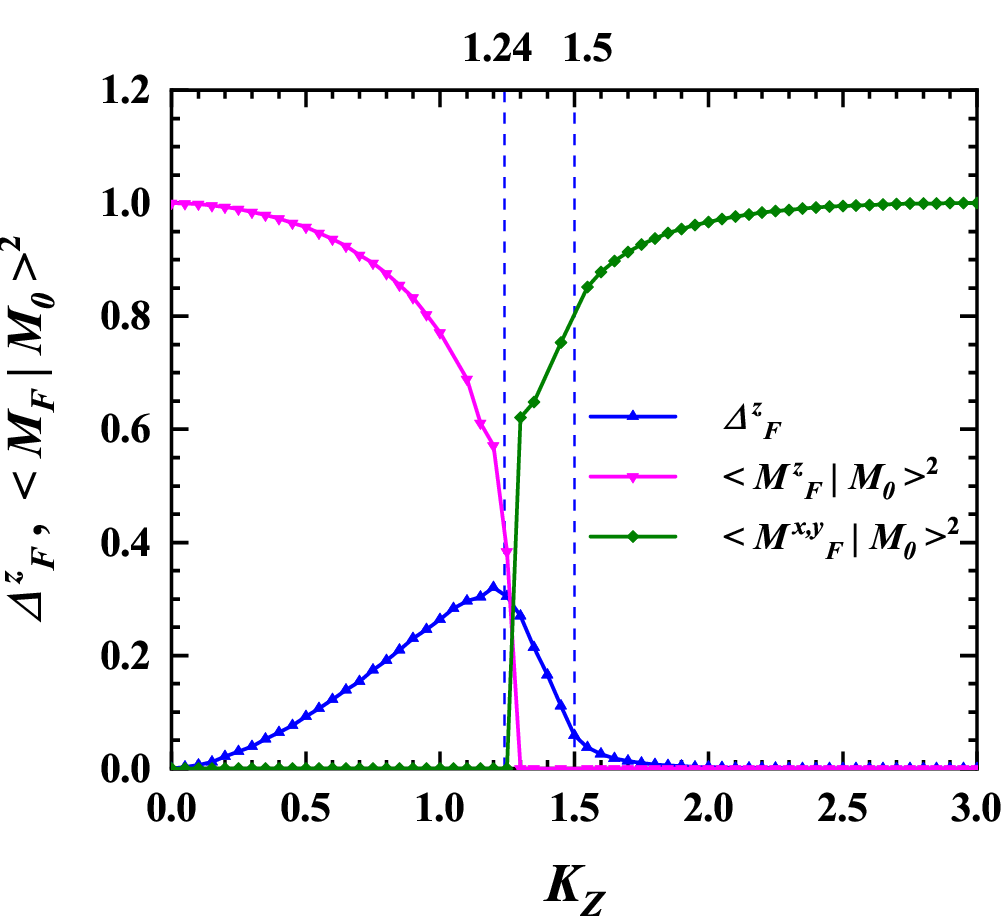}}
\hspace{0in}
\subfigure[]{
\label{fig:state3c}
\includegraphics[angle=0, width=0.66 \columnwidth]{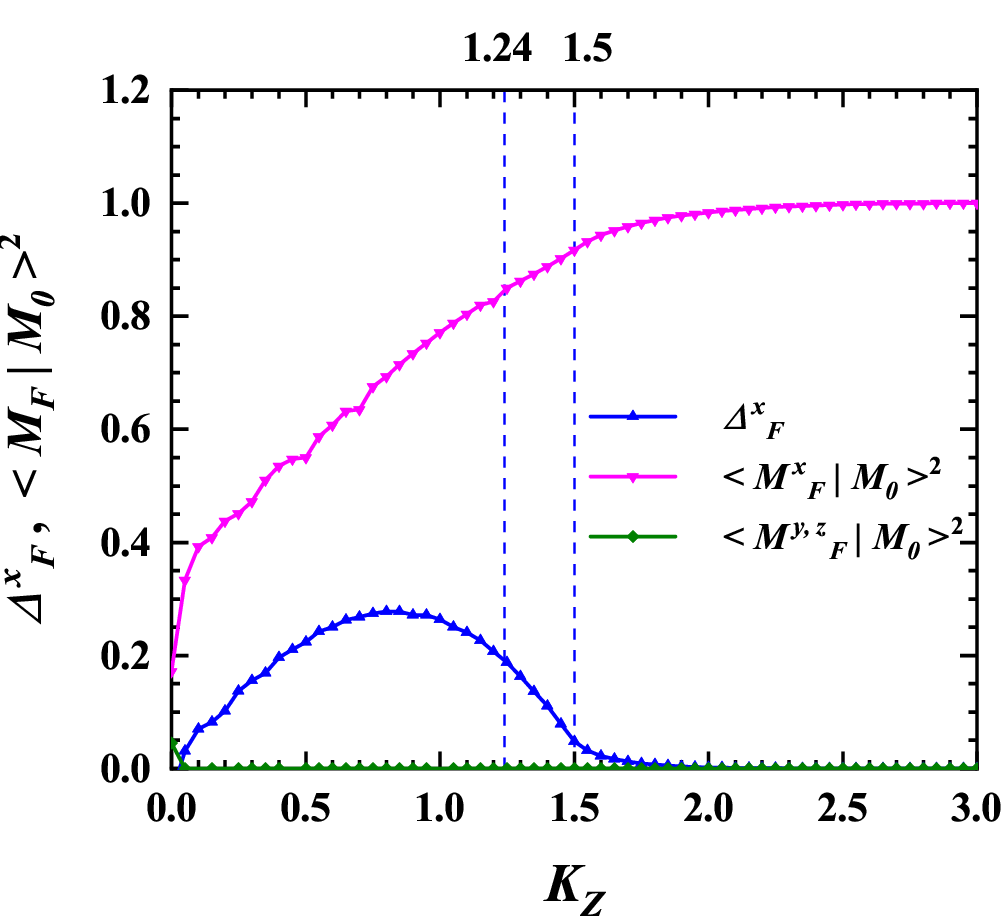}}
\caption{(Color online) (a) The variation range of Kitaev coupling strengths in the parametric phase diagram of the
 Kitaev model with the conditions $K_{X}=K_{Y}$ and $K_{X}+K_{Y}+K_{Z}=3.0$, marked by red, blue and green arrows. Six points are marked with black dots, $K_{Z}=1.8, 1.5, 1.4, 1.24, 1.0$, and $0.6$. Kitaev coupling $K_{Z}$ dependences of (b) the overlaps $\langle M_{F}^{z}|M_{0}\rangle^2$, $\langle M_{F}^{x,y}|M_{0}\rangle^2$ and vison gap $\Delta_{F}^{z}$, and (c) the overlaps $\langle M_{F}^{x}|M_{0}\rangle^2$, $\langle M_{F}^{y,z}|M_{0}\rangle^2$ and vison gap $\Delta_{F}^{x}$ in the variation range of (a).}
\label{fig:state3}
\end{figure*}

Through the diagonalization of the zero-flux Hamiltonian matrix in the momentum space, the ground-state spinon energy dispersion can be expressed as
\begin{eqnarray}
 \label{eq:Hamiltonian12}
  E_{\mathbf{k}}=2\left|K_{X}e^{i\mathbf{k\cdot X}}+K_{Y}e^{i\mathbf{k\cdot Y}}+K_{Z}e^{i\mathbf{k\cdot Z}}\right|
 \end{eqnarray}
The ground-state parametric phase diagram is obtained \cite{Kitaev2006}, as shown in Fig.~\ref{fig:state3}(a). From Eq.~(\ref{eq:Hamiltonian12}), one can find a van Hove singularity at $E_{V1}=2|K_{Z}|$ corresponding to the energy contour line PMP' in the first Brillouin region; and another van Hove singularity at $E_{V2}=2|K_{X}+K_{Y}-K_{Z}|$ when $|K_{Z}|<1.5$, or a spinon gap $\Delta_{S}=2|K_{Z}-K_{X}-K_{Y}|$ when $1.5<|K_{Z}|<3.0$, associated with the M' point. There is also a energy maximum $E_{max}=2|K_{X}+K_{Y}+K_{Z}|=6$ at $\Gamma$ point.
In probing into the dynamical features and evolution of the QSL ground states in the anisotropic Kitaev model, we take the range of the Kitaev couplings $K_Z$ along the line marked by red, blue, and green lines with arrows, labelling the gapped, gapless and another gapless QSL, in this phase diagram. The quantum phases in these three regions display distinct different quantum features \cite{Jia2021}.

The time-dependent dynamical spin susceptibility of the ground state, $S_{ij}^{\alpha\alpha}(t)=0.25\langle\Phi_{0}|\hat{\sigma}_{i}^{\alpha}(t)\hat{\sigma}_{j}^{\alpha}(0)|\Phi_{0}\rangle$ ($|\Phi_{0}\rangle=|F_{0}\rangle \otimes |M_{0}\rangle$) \cite{Knolle2016}, can be derived as follows:
\begin{eqnarray}
\label{eq:Hamiltonian13}
 & & S_{ij}^{\alpha\alpha}\left(t\right) = -0.25i\left\langle M_{0}\left|e^{iH_{0}t}c_{i}
 e^{-i\left(H_{0} + V_{\langle ij\rangle_{\Lambda}}\right)t}c_{j}\right|M_{0}\right\rangle \nonumber\\
 & & \left(i\delta_{ij} + \hat{u}_{ij}^{\alpha}\delta_{\langle ij\rangle,\Lambda}\right),
\end{eqnarray}
where $V_{\langle ij\rangle_{\Lambda}} = -2iK_{\Lambda}c_{i}c_{j}$, $i\in A,j\in B$, and $\alpha = x,y,z$ corresponds to $\Lambda=X,Y,Z$ one-to-one. We can find that only the on-site ($\delta_{ij}$) and NN ($\delta_{\langle ij\rangle,\Lambda}$) ones of the dynamical spin correlation are non-zero, and $S_{ij}^{\alpha\alpha}$ only has the $\alpha=z(x,y)$ component in the NN $X(Y,Z)$-bond.

$S_{ij}^{\alpha\alpha}$ has the {\it Lehmann} representation by inserting the identity $\mathbf{1}=\sum_{\lambda}|\lambda\rangle\langle\lambda|$ of the two-flux sector with a flipping bond $u_{ij}^{\alpha}=-1$. The main contributions are from the zero-, one- and two-particle of $|\lambda\rangle$, which occupy the $98\%$ of the total \cite{Knolle2016}. We thus can obtain the dynamical spectrums in the frequency $\omega$-space as
\begin{eqnarray}
\label{eq:Hamiltonian14}
 &S_{AA}^{\alpha\alpha}\left(\omega\right)=\frac{\pi}{2}\sum\limits_{\lambda}{\left\langle{M_{0}}\right|{c_{A}}\left|\lambda  \right\rangle \left\langle\lambda\right|{{c}_{A}}\left|{{M}_{0}}\right\rangle \delta[\omega -\left(E_{\lambda }^{F}-{{E}_{0}}\right)]}, \nonumber \\
 &S_{BA}^{\alpha\alpha}\left(\omega \right)=\frac{\pi}{2} i\sum\limits_{\lambda}{\left\langle  {{M}_{0}}\right|{{c}_{B}}\left|\lambda  \right\rangle \left\langle\lambda\right|{{c}_{A}}\left| {{M}_{0}}\right\rangle \delta[\omega -\left(E_{\lambda }^{F}-{{E}_{0}}\right)]}. \nonumber \\
\end{eqnarray}
Here $E_{0}$ is the ground-state energy of the zero-flux sector, and $E_{\lambda}^{F}$ is the energy eigenvalue of the \cite{Knolle2016}state $|\lambda\rangle$ of two-flux sector, while the lowest-energy is $E_{0}^{F}$ with the state $|M_{F}^{z(x,y)}\rangle$.  $|\lambda\rangle$ and $E_{\lambda}^{F}$  are obtained through the diagonalization of the two-flux Hamiltonian matrix in the real space. Further we can calculate the overlaps $\langle M_{0}|M_{F}^{z(x,y)}\rangle^2$ and ``vison'' gap $\Delta_{F}^{z(x,y)}=E_{0}^{F}-E_{0}$ due to the gauge-flux excitation, as shown in Fig.~\ref{fig:state3}(b)(c), consistent with Knolle's results \cite{Knolle2016}.

From the dynamical phase diagrams in Fig.~\ref{fig:state3}(b)(c), we can see that the lowest-energy states of the zero-flux sector $H_{0}$ and two-flux sector $H_{0}+V_{z(x,y)}$, $|M_{0}\rangle$ and $|M_{F}^{z(x,y)}\rangle$ conserve the parity owing to the spatial inversion symmetry. Along the line in Fig.~\ref{fig:state3}(a), $|M_{0}\rangle$ and $|M_{F}^{z}\rangle$ have the same parity when $1.24<|K_{Z}|<3.0$ and the opposite parity when $0<|K_{Z}|<1.24$; $|M_{0}\rangle$ and $|M_{F}^{x(y)}\rangle$ have the same parity  all the way.
In the case with the same parity, $|\lambda\rangle$ must contain the odd number of excitations, mainly the single-particle contribution. This dynamical spin susceptibilities could be evaluated by Eq.(12). In the opposite case, $|\lambda\rangle$ must contain the even number of excitations, mainly the zero- and two-particle contributions. Actually the {\it Lehmann} representation is modified by inserting the identity $1=\sum_{\lambda}c_{A(B)}|\lambda\rangle\langle\lambda|c_{A(B)}$ of two-flux sector $H_{0}+V_{x}+V_{y}$ with two flipping bonds $u_{ij}^{x},u_{ij}^{y}=-1$. Its lowest-energy state $|M_{F}^{x,y}\rangle$ have the same parity with $|M_{0}\rangle$ \cite{Knolle2016}, as shown in Fig.~\ref{fig:state3}(b). We also plot the lowest-energy state $|M_{F}^{y,z}\rangle$ for $H_{0}+V_{y}+V_{z}$ shown in Fig.~\ref{fig:state3}(c). It has the opposite parity with $|M_{0}\rangle$ all the way.
Therefore, we can explicitly express Eq.~(\ref{eq:Hamiltonian11}) for the zero- and two-particle contributions,
\begin{eqnarray}
\label{eq:Hamiltonian15}
 & S_{AA}^{\alpha\alpha}\left(\omega\right)=\frac{\pi}{2}\sum\limits_{\lambda}{\left\langle{M_{0}}|\lambda\right\rangle \left\langle\lambda|{M_{0}}\right\rangle \delta \left[\omega-\left(E_{\lambda }^{F}-{{E}_{0}} \right) \right]}, \\
 & S_{BA}^{\alpha\alpha}\left(\omega \right)=\frac{\pi}{2} i\sum\limits_{\lambda}{\left\langle{M_{0}} \right|{c_{B}}{c_{A}}\left|\lambda\right\rangle \left\langle\lambda|{M_{0}}\right\rangle
 \delta\left[\omega-\left(E_{\lambda}^{F}-{E_{0}} \right) \right]}. \nonumber
\end{eqnarray}
Then the dynamical spin correlation $S^{m}(\epsilon)=S^{m}_{AA}(\epsilon)+S^{m}_{BA}(\epsilon)$ ($m=\alpha\alpha$,$\alpha=x,y,z$). More details are shown in Sec. C of {\it Supplementary Materials} \cite{Supple2020}.
Hence, combining the dynamical and parametric phase diagrams, we choose four representative points $K_{Z}=1.8, 1.4, 1.0$ and $0.6$, respectively, among the phase transition points about the parity relationship and spinon gap, $K_{Z}=1.24$ and $1.5$.


Substituting the Eq.~(\ref{eq:Hamiltonian8}),(\ref{eq:Hamiltonian14}) and (\ref{eq:Hamiltonian15}) into Eq.~(\ref{eq:Hamiltonian6}), (\ref{eq:Hamiltonian9}) and (\ref{eq:Hamiltonian10}), we can obtain the tunneling current $I^{c,s}$ and differential conductances $dI^{c}/dV$ and $dI^{s}/d\Delta$.
Throughout this {\it paper} the SC order parameters $\tilde{\Delta}_{L}$ and $\tilde{\Delta}_{R}$ in the left and right leads have the same modulus $\Delta_{L}=\Delta_{R}=\Delta$, but different phase $\phi_{L(R)}$.
In this paper, all of the energies are measured in terms of the Kitaev coupling $K$, which can be taken as $K=1$.

\section{\label{sec:level3} Results and Discussion}
\subsection{\label{sec:level31} Dynamical spin correlations of the anisotropic Kitaev model}

\begin{figure*}[htbp]
\centering
\includegraphics[angle=0, width=2.0 \columnwidth]{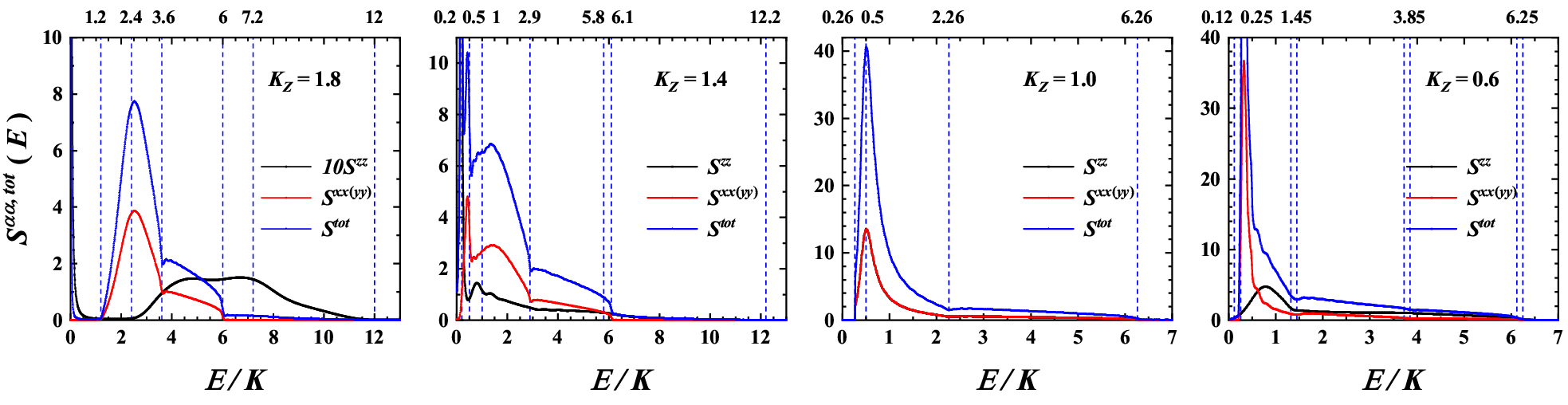}
\caption{(Color online) Energy $E$ dependences of the dynamical spin susceptibilities of the anisotropic Kitaev QSL, including the total  $S^{tot}$ and its three components $S^{\alpha\alpha} (q=0,E) (\alpha=x,y,z)$ for different Kitaev coupling $K_{Z}=1.8$ (a), $1.4$ (b), $1.0$ (c) and $0.6$ (d), respectively, in units of the energy $K$. Here $S^{xx}=S^{yy}$}
\label{fig:state4}
\end{figure*}

At first, we plot the dynamical spin susceptibilities of the anisotropic Kitaev model, including the components $S^{\alpha\alpha}(E) (\alpha=x,y,z)$ and their total $S^{tot}$, as functions of energy $E$ \cite{Carrega2020,Konig2020,Feldmeier2020,Knolle2014,Knolle2016}, as shown in Fig.~\ref{fig:state4}(a)-(d). Here $S^{xx}=S^{yy}$ because $K_{X}=K_{Y}$.
From this, we can see that the anisotropic components of dynamical spin susceptibilities, $S^{zz}$ and $S^{xx(yy)}$, and their total $S^{tot}$ reveal remarkable different features in these four quantum phases.

When $K_{Z}=1.8$ with a gapped QSL, the parities between $|M_{F}^{z}\rangle$ and $|M_{0}\rangle$ are opposite. As shown in Fig.~\ref{fig:state4}(a), in $S^{xx(yy)}$, we can see the total QSL gap $\Delta_{t}\approx 1.2$. It actually contains the spinon gap $\Delta_{S}=2|K_{Z}-K_{X}-K_{Y}|=1.2$ and vison gap $\Delta_{F}^{x}\approx0.0$. There is a dip at $E\approx3.6$ owing to the van Hove singularity of spinon spectrum at $2K_{Z}$ and an energy shift of $\Delta_{F}^{x}$. And an upper edge emerges at about $6.0$ which equals $\Delta_{F}^{x}+2|K_{X}+K_{Y}+K_{Z}|$.
So these three feature points in $S^{xx(yy)}$ correspond to the ones of spinon dispersion at $E_{V}$ ($\Delta_{S}$ or $E_{V1}$, $E_{V2}$ and $E_{max}$) one-to-one, and move towards $\Delta_{F}^{x(y)}+E_{V}$.
There is a new peak at about $2.5$ caused by the interacting vison and spinon.
However, in $S^{zz}$, we can observe the total gap $\Delta'_{t}\approx 2.4$, which stems from the $\Delta_{F}^{z}\approx0.0$ and the new spinon gap $\Delta'_{S}=2\Delta_{S}$. A peak appears at $E\approx7.2$ resulted from the van Hove singularity, and the upper edge emerges at $E\approx12.0$. The three feature points at $\Delta_{F}^{z}+2E_{V}$ in $S^{zz}$ are from the virtual transitions to the eigenstates of two-flux sector with two flipping bonds.
Moreover, we can see a distinct sharp peak at $\Delta_{F}^{z}$, stemed from the virtual transitions to the lowest-energy state, $|M_{F}^{x,y}\rangle$. There is also a new peak around $5.0$ due to the interaction of vison and spinon. Note that $S^{xx(yy)}$ is an order of magnitude bigger than $S^{zz}$.
As for the sum of $S^{zz}$, $S^{xx}$ and $S^{yy}$, $S^{tot}$ can exhibit the complete information of vison, spinon and their interaction, except some feature points because of the resolution of $S^{zz}$.

When $K_{Z}=1.4$ shown in Fig.~\ref{fig:state4}(b), the ground-state is gapless QSL and $|M_{F}^{z}\rangle$ and $|M_{0}\rangle$ have the opposite parity. Hence, the three feature points are displayed on $S^{xx(yy)}$ and $S^{zz}$ in the similar way as $K_{Z}=1.8$, except the van Hove singularity instead of the spinon gap. In $S^{xx}$, we can observe two dips at $E\approx0.5$ and $2.9$ corresponding to the van Hove singularities, and a upper edge at about $6.1$ with $\Delta_{F}^{x}\approx0.11$. These three feature points emerge at $\Delta_{F}^{x(y)}+E_{V}$. There are two new interaction peaks at about $0.4$ and $1.5$.
In $S^{zz}$, there is a dip and an inflection point associated with the van Hove singularities at $E\approx1.0$ and $5.8$, and a boundary at about $12.2$ with $\Delta_{F}^{z}\approx0.17$. So these feature points are shown at $\Delta_{F}^{z}+2E_{V}$. A remarkable sharp peak appears at $\Delta_{F}^{z}$, and a new interaction peak emerges at about $0.8$.
Since $S^{zz}$ has the same order in magnitude to $S^{xx,yy}$, the total one $S^{tot}$ could reveal the full dynamical features of Kitaev QSL well.

When $K_{Z}=1.0$ and $0.6$, as shown in Fig.~\ref{fig:state4}(c)(d), $|M_{F}^{z(x,y)}\rangle$ and $|M_{0}\rangle$ have the same parity.
At $K_{Z}=1.0$, the ground-state of the isotropic Kitaev model is a $C_{6}$ gapless QSL, and $S^{xx(yy)}$ and $S^{zz}$ components are equal, $\Delta_{F}^{x(y)}=\Delta_{F}^{z}\approx0.26$. From $S^{zz}$ we can find that there is only one dip related to the two-in-one van Hove singularity at $E\approx2.26$, and a upper edge at about $6.26$. There is also a new interaction peak at about $0.5$.
At $K_{Z}=0.6$, two dips resulted from the van Hove singularities, a upper edge and a new sharp peak are shown at $E\approx1.45,3.85$, $6.25$ and $0.3$ in $S^{xx(yy)}$ with $\Delta_{F}^{x(y)}\approx0.25$. There is a dip and a inflection point, a upper boundary and a new peak at $E\approx1.32,3.72$, $6.12$ and $0.8$ in $S^{zz}$ with $\Delta_{F}^{z}\approx0.12$.
Therefore, the feature points emerge at $\Delta_{F}^{x(y)}+E_{V}$ and $\Delta_{F}^{z}+E_{V}$ in $S^{xx(yy)}$ and $S^{zz}$, respectively, because of the virtual transitions to the eigenstates of two-flux sector with only a flipping $X(Y)$- or $Z$-bond. The total spin correlations $S^{tot}$ also have the entire characters of Kitaev QSL when $K_{Z}=1.0$ and $0.6$.

In a word, the dynamical spin susceptibility components $S^{zz}$ and $S^{xx}(=S^{yy})$ reveal different vison gaps $\Delta_{F}^{z}$ and $\Delta_{F}^{x(y)}$, respectively. Every component can only reveal the partial features of the Majorana fermion (spinon) dispersions influenced by the gauge fluxes, including the two van Hove singularities (or a spinon gap and a van Hove singularity) and the energy upper edge, and in different ways. Therefore, the total $S^{tot}$ can exhibit the complete information of Kitaev QSL well.
There are some new peaks between these feature points, which stem from the interaction between the vison and spinon excitations.

\subsection{\label{sec:level32} DC Josephson current with zero voltage}

\begin{figure*}[htbp]
\centering
\includegraphics[angle=0, width=2.0 \columnwidth]{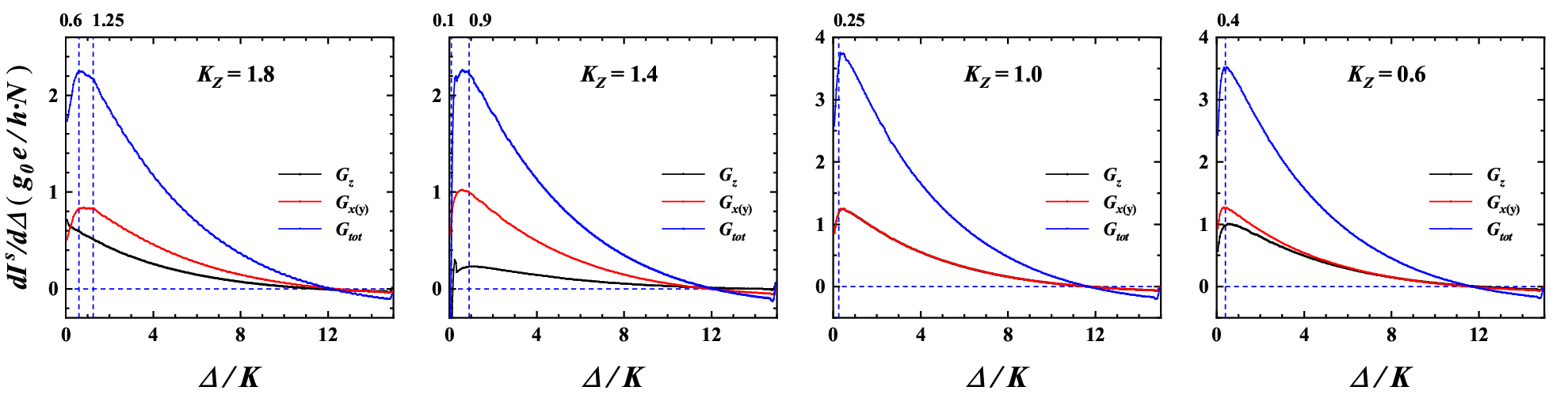}
\caption{(Color online) Derivative of the DC {\it Josephson} tunneling current $I^{s}$ with respect to the SC gap, $dI^{s}/d\Delta$, including the components $G_{z(x,y)}$ ($G_{x}=G_{y}$) and the total $G_{tot}$, as functions of the SC gap $\Delta$ for different Kitaev coupling $K_{Z}=1.8$ (a), $1.4$ (b), $1.0$ (c) and $0.6$ (d), respectively. Here $\phi=3\pi/2$}
\label{fig:state5}
\end{figure*}

In the absence of the bias voltage, only the DC {\it Josephson} current with the tunneling of the Cooper pairs is presented in the SC-Kitaev QSL-SC junction. The SC gap $\Delta$ dependences of the derivative of the DC {\it Josephson} current $I^{s}$ with respect to $\Delta$, $G_{tot}=dI^{s}/d\Delta$, and its components $G_{z(x,y)}$ ($G_{x}=G_{y}$) have been described in Fig.~\ref{fig:state5}(a)-(d) for different Kitaev couplings $K_{Z}=1.8,1.4,1.0$ and $0.6$, respectively. Here the phase difference $\phi=3\pi/2$ and we define the dimensionless constant $g_{0}=4\pi\rho_{L}\rho_{R}J^{2}$.

As seen in Fig.~\ref{fig:state5}(a)-(d), when $K_{Z}=1.8$, we can see a peak at $\Delta\approx0.6$, and an inflection point at about $1.25$. The former corresponds to the total QSL gap $\Delta_{t}\approx1.2$, which originates from the resonant tunneling when $2\Delta=\Delta_{t}$, while the latter stems from the interaction between the vison and spinon excitations near $2\Delta\approx2.5$.
Similarly, at $K_{Z}=1.4$, a distinct peak, corresponding to the total QSL gap, emerges around $2\Delta\approx0.17$. Another peak at about $0.9$ stemming from the response to the interaction peak appears near $2\Delta\approx1.8$.
When $K_{Z}=1.0$ and $0.6$, we can only observe the peaks at about $0.25$ and $0.4$, which are due to the response to the interaction peaks around $2\Delta\approx0.5$ and $0.8$, respectively. Thus, $dI^{s}/d\Delta$ curves mainly provide the information about the interaction of gauge fluxes and Majorana fermion mode, as well as the total QSL gap.
These peaks in $dI^{s}/d\Delta$ could be seen from the dynamical spin susceptibilities in Fig.~\ref{fig:state4}, however, only partially provide full information of the Kitaev QSLs.

\begin{figure*}[htbp]
\centering
\includegraphics[angle=0, width=2.0 \columnwidth]{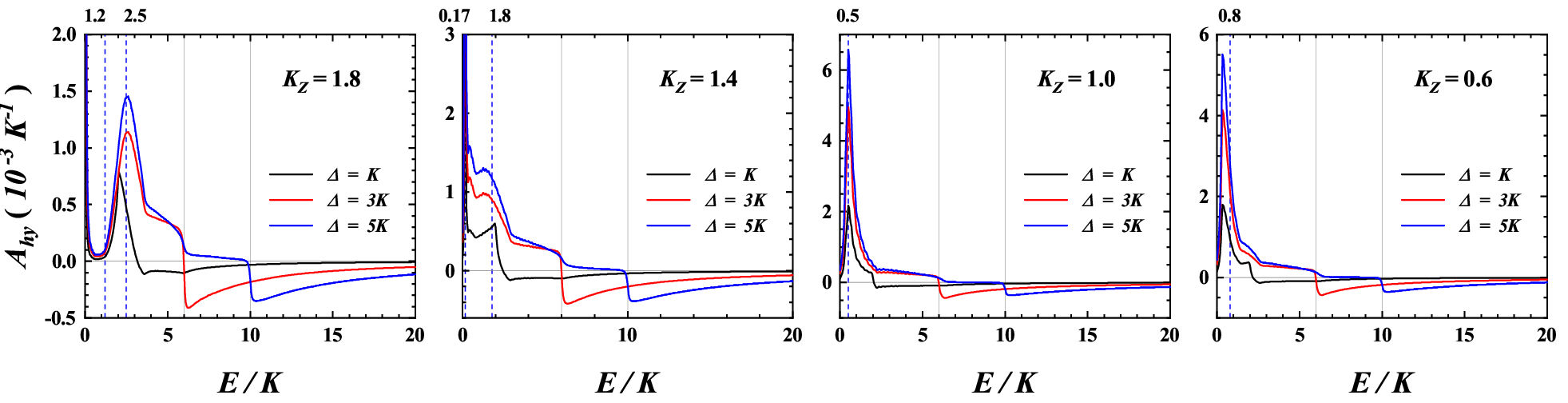}
\caption{(Color online) Energy $E$ dependences of the total hybridization spectral functions between spins of the central Kitaev QSL layer and Cooper pairs of the two SC leads, $A_{hy}$, with $\Delta$=1, 3 and 5 for different Kitaev coupling $K_{Z}=1.8$ (a), $1.4$ (b), $1.0$ (c) and $0.6$ (d), respectively.}
\label{fig:state6}
\end{figure*}

To understand this reason, we plot the energy $E$ dependences of the total dynamical hybridization spectral functions between local spins of the Kitaev layer and Cooper pairs of the two SC leads, $A_{hy}=\sum_{\alpha}A_{hy}^{\alpha\alpha}$ ($\alpha=x,y,z$), in Fig.~\ref{fig:state6}(a)-(d). The SC gaps are set as $\Delta$=1, 3, and 5, respectively.
From Fig.~\ref{fig:state6} we can see the whole dynamical spin correlation characters clearly and $A_{hy}>0$ before $E=2\Delta$. When $E>2\Delta$, these spin correlation features appear with a reversal sign, {\it i.e.} $A_{hy}<0$ in the same magnitude. Thus, the DC {\it Josephson} current at zero bias, as the frequency integration of the hybridization spectrum, is partially cancelled; hence, it only keeps partial information of Kitaev QSL.
This arises from the fact that in the inelastic tunneling, the quasi-electrons and quasi-holes of the SC Cooper pairs contribute the positive and negative parts of the $A_{hy}$, respectively. Therefore, the total response to the dynamical spin correlation spectrum is cancelled out due to the spin-singlet Cooper pairs.

\subsection{\label{sec:level33} DC conductance of the normal single-particle tunneling}

\begin{figure*}[htbp]
\centering
\includegraphics[angle=0, width=2.0 \columnwidth]{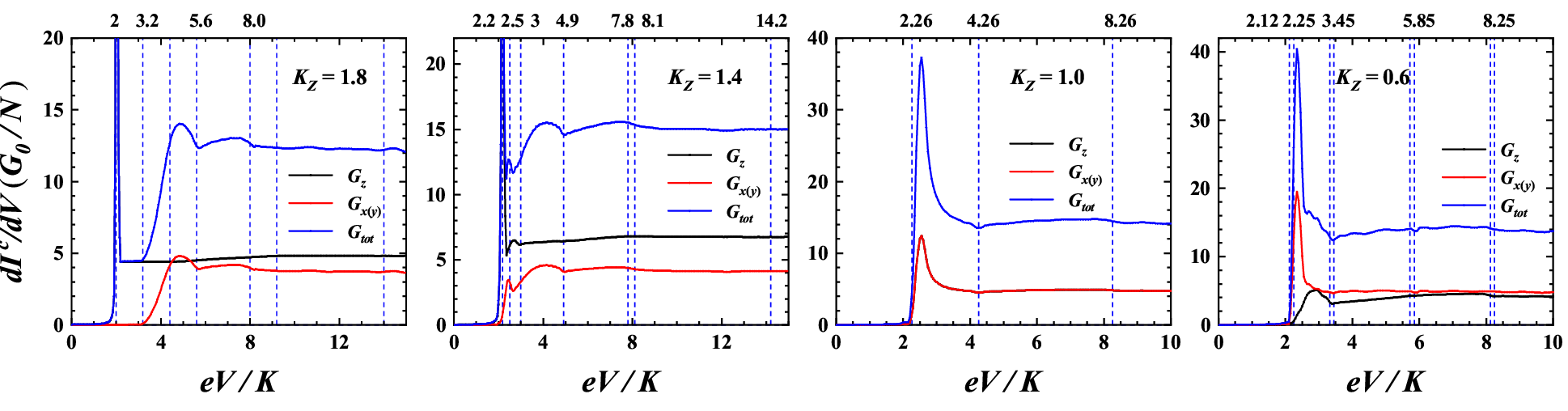}
\caption{(Color online) DC differential conductances of the single-particle tunneling $dI^{c}/dV$, including the components $G_{z(x,y)}$ ($G_{x}=G_{y}$) and the total $G_{tot}$, as functions of the bias potential $eV$ for different Kitaev coupling $K_{Z}=1.8$ (a), $1.4$ (b), $1.0$ (c) and $0.6$ (d), respectively.}
\label{fig:state7}
\end{figure*}

Further, in the presence of a DC bias voltage in the SC-Kitaev QSL-SC junction, one could reveal more characters of the Kitaev QSL. The bias potential $eV$ dependences of the DC single-particle differential conductance  $G_{tot}$=$dI^{c}/dV$, as well as its $zz(xx,yy)$ components $G_{z(x,y)}$, have been described in Fig.~\ref{fig:state7}(a)-(d) for different Kitaev couplings $K_{Z}=1.8,1.4,1.0$ and $0.6$, respectively. Here we define the conductance constant $G_{0}=g_{0}e^{2}/h$, and $G_{tot}=G_{x}+G_{y}+G_{z}$ with $G_{x}=G_{y}$.

From Fig.~\ref{fig:state7}(a)-(d), the single-particle DC differential conductance spectrums of the SC junction, $G_{tot}$ and $G_{z(x,y)}$,  show the distinct different characters in the four quantum phases. To clearly see the dynamical behaviors of $G_{tot}$ in present anisotropic Kitaev layer, we first describe the bias voltage dependence of $G_{z(x,y)}$.
When $K_{Z}=1.8$, as seen in Fig.~\ref{fig:state7}(a), contrast with the dynamical spin susceptibility in Fig.~\ref{fig:state4}(a), the threshold of the conductance $G_{x(y)}$ is modulated up to about $3.2$, {\it i.e.} $\Delta_{t}+2\Delta$. This arises from the fact that the electrons at the bottom of the SC gap in the right lead need a high enough bias potential $eV=\Delta+\Delta+\Delta_{t}$ to overcome the right and left SC gaps and total QSL gap of the central layer along the $X$-($Y$-) bond, and finally reach the empty state on the top of the SC gap in the left lead.
When $eV > 2\Delta+\Delta_{t}$, with the open of the channel of the Majorana bond state, the conductance $G_{x(y)}$ starts to rise rapidly and goes up to a sharp peak at about $4.5$. This peak corresponds to the interaction peak of the dynamical spin correlation around $2.5$ shown in Fig.~\ref{fig:state4}(a), and results from the dynamical creation of the Majorana fermions (or spinon) interacting with the NN two gauge fluxes in the virtual transition.
Soon afterwards, a remarkable dip have been observed at about $5.6$, which is associated with the dip of dynamical spectrum around $3.6$ and due to the van Hove singularities of the DOS of the Majorana dispersive band. Finally, the single-particle conductance approaches to a constant after the upper edge at about $8$ due to the one of Majorana dispersive band around $6$. Hence, the features of the single-particle DC differential conductance spectrums $G_{x(y)}$ in Fig.~\ref{fig:state7}(a)correspond to those of the dynamical spin susceptibility one-to-one in Fig.~\ref{fig:state4}(a).

Meanwhile, we can see a remarkable sharp peak at $eV\approx 2$ in $G_{z}$ related to the one in the dynamical spectrum near $\Delta^{F}_{z}\approx0.0$, which originates from the $\delta$-function contribution of the virtual transition between the ground state $|M_{0}\rangle$ and the excited state $|M_{F}^{x,y}\rangle$. When $eV > \Delta^{F}_{z}$, no obvious characters in $G_{z}$ is observed since $S^{zz}$ is an order of magnitude smaller than $S^{xx(yy)}$.
Summing the three components gives rise to the total conductance $G_{tot}$, which contains the complete characters of the spinon spectrums, vison excitation and their interaction of Kitaev QSL.
Thus, compared to the normal-metal junction situation\cite{Carrega2020,Konig2020}, the present differential conductance spectrums $G_{tot}$ have a more intuitive and sensitive response to the characters of dynamical spin correlation components of Kitaev QSL, $S^{tot}$.

When $K_{Z}=1.4, 1.0$ and $0.6$, similar to $K_{Z}=1.8$, the single-particle DC differential conductance spectrums $G_{z(x,y)}$ can reflect the features of dynamical spin susceptibility components $S^{zz(xx,yy)}$ well, except some feature points due to the resolution in numerical integration. Fortunately, in the present situations with $K_{Z}=1.4, 1.0$ and $0.6$, the z-component of dynamical spin correlations $S^{zz}$ are the same order in magnitude to $S^{xx,yy}$, so $G_{z}$ can resolve the complete features of $S^{zz}$.
Hence, from the single-particle tunneling spectrums, we could get insight into the features of the dynamical spin susceptibilities of Kitaev QSL.

\section{\label{sec:level4} Conclusion}

In our present theory, we point out two possible improvements to the present results. On the one hand, with the condition of $\mathbf{q \cdot X(Y,Z)}\approx0$, we obtain the features of the total dynamical spin susceptibility $S^{tot}$. When $\mathbf{q \cdot X(Y,Z)}\neq0$, the individual contribution of each component of the NN spin correlation $S^{\alpha\alpha}_{BA}$ to the tunneling currents would be slightly different from the result above. Our further study reveals that in this situation the correction to Eq.~(\ref{eq:Hamiltonian8}) only quantitatively alters the tunneling current, nevertheless, it is qualitatively consistent with the above conclusion. On the other hand,  although the zero-voltage {\it Josephson} current fails to measure the full information of the Kitaev QSL in the elastic scattering process, we expect that the AC {\it Josephson} currents with DC bias voltage can reveal more features of dynamical spin correlation, which goes beyond the scope of this paper.

As a summary, in investigating the electron tunneling transport and its spectroscopic features in an SC-anisotropic Kitaev QSL-SC {\it Josephson} junction with the weak link, we assume that the inelastic scattering tunneling of the single particle and Cooper pair is realized by the $s-d$ exchange interaction.
As expected, the DC differential conductance $dI^{c}/dV$ of the normal single-particle tunneling succeeds in exhibiting the dynamical spin susceptibility characters of the anisotropic Kitaev QSL, including the unique spin gaps even in gapless QSL, the sharp or broad peaks, the small dips and the upper edge of the itinerant Majorana fermion dynamics, except an energy shift of two-SC-lead gap $2\Delta$. The different topological quantum phases of anisotropic Kitaev QSL can be distinguished by the tunneling spectral features well.
Unusually, the zero-voltage DC {\it Josephson} currents $I^{s}$ only have some residual information of Kitaev QSL, which stems from the spin singlet of Cooper pairs. Our results may pave a new way to measure the Majorana-fermion dynamical correlation features of the anisotropic Kitaev and other spin liquid materials. We expect that our theoretical results could be confirmed by future experiments and be applied in the SC junction devices.

\begin{acknowledgments}
L. J. thanks the supports from the NSFC of China under Grant Nos.11774350 and 11474287, H.Q. acknowledges financial support from NSAF U1930402 and NSFC 11734002. Numerical calculations were performed at the Center for Computational Science of CASHIPS and Tianhe II of CSRC.
\end{acknowledgments}

\providecommand{\noopsort}[1]{}\providecommand{\singleletter}[1]{#1}%
%



\end{document}